\documentclass[twoside,11pt]{article}
\usepackage[tbtags]{amsmath}
\usepackage{amssymb}
\usepackage{dsfont}
\usepackage{a4wide}
\usepackage{fancyhdr}
\usepackage{latexsym}
\numberwithin{equation}{section}
\makeatletter
\title{The partition function of the linear Poisson-sigma model on arbitrary surfaces}
\author{ Allen C. Hirshfeld\footnote{\tt hirsh@physik.uni-dortmund.de}$\;\;$ and Thomas 
Schwarzweller\footnote{\tt thomas@zylon.physik.uni-dortmund.de}\\Fachbereich Physik\\Universit\"at Dortmund}
\begin{document}
%
%%%%%%%%%%%%%%%%%%%%%%%%%%%%%%%%%%%%%%%%%%%%%%%%%%%%%%%%%%%%%%%%%%%%%%%%%%%%%%%%%%%%%%%%%
%
\newcommand{\sss}{\scriptscriptstyle}
\newcommand{\mc}{\mathcal}
\newcommand{\bmult}{\begin{multline}}
\newcommand{\emult}{\end{multline}}
\newcommand{\rd}{\mathrm d}
\newcommand{\Vol}{\text{Vol}}
%
%%%%%%%%%%%%%%%%%%%%%%%%%%%%%%%%%%%%%%%%%%%%%%%%%%%%%%%%%%%%%%%%%%%%%%%%%%%%%%%%%%%%%%%%%
%
\pagestyle{fancy}
\lhead{}
\chead{\sl\small A.C. Hirshfeld and T. Schwarzweller, The linear Poisson-sigma model on arbitrary surfaces}
\rhead{}
\renewcommand{\headrulewidth}{0.3pt}
\renewcommand{\footrulewidth}{0pt}
%
%%%%%%%%%%%%%%%%%%%%%%%%%%%%%%%%%%%%%%%%%%%%%%%%%%%%%%%%%%%%%%%%%%%%%%%%%%%%%%%%%%%%%%%%%
%
\maketitle
\begin{abstract}
We perform the calculation of the  partition function of the Poisson-sigma model on the world sheet with 
the topology of a two-dimensional disc. Considering the special case of a linear Poisson structure we recover the partition function of the Yang-Mills theory. Using a glueing procedure we are able to 
calculate the partition function for arbitrary base manifolds.
\end{abstract}
%
%
%%%%%%%%%%%%%%%%%%%%%%%%%%%%%%%%%%%%%%%%%%%%%%%%%%%%%%%%%%%%%%%%%%%%%%%%%%%%%%%%%%%%%%%%%
%
\section{Introduction}
The Poisson-sigma model has attracted increasing interest in recent years. Originally investigated  by Schaller and Strobl as a 
generalization of 2d gravity-Yang-Mills systems \cite{SS} and independently by Ikeda as a non-linear gauge theory \cite{I} it turned out 
to be more than a unified  treatment of these specific models. 
Actually, the Poisson-sigma model  associates to various Poisson structures on finite dimensional manifolds two-dimensional 
field theories which include gravity models \cite{TS,KS,EKS}  and non-Abelian gauge theories, in particular Yang-Mills theory, and the 
gauged Wess-Zumino-Witten model \cite{ASS}. 
It was already noted in \cite{SS,SS2} that the quantization of the Poisson-sigma model as a field theory
has implications for more general questions concerning the quantization of various spaces, in particular
the quantization of Poisson 
manifolds. By use of the canonical quantization procedure it was shown that the symplectic leaves of the target manifold must satisfy 
an integrality condition. In the meantime 
Catteneo and Felder \cite{CF1} have shown that the perturbation expansion of the path integral in the covariant gauge 
reproduces the Kontsevich's formula for the deformation quantization of the algebra of functions on a Poisson manifold \cite{K}. 

The connection to gravity models was used by Kummer, Liebl and Vassilevich  to investigate the special case of 2d dilaton gravity in 
the temporal gauge, and they 
have calculated the generating functional using BRST methods \cite{KLV1}. In further work they have studied the coupling to matter 
fields \cite{KLV2}. 

In \cite{HS} we have used path integral techniques to derive a general expression for the partition function of the Poisson-sigma model 
on closed manifolds for an arbitrary gauge.  In this calculation we reproduce the quantization condition for the symplectic leaves to 
be integral, now for arbitrary closed world sheets. Further, we have shown that for a linear Poisson structure the partition function is 
fully computable and 
the partition function for the Yang-Mills theory may be recovered from that of the linear Poisson-sigma model. 
Klimcik \cite{KC} has introduced a more general model where the target spaces are given by certain  {\it so-called} Drinfeld doubles \cite{D}, 
such that the Poisson-sigma model with a Poisson-Lie group as the target space is included. He calculates the partition function for this model, which turns out to be 
a q-deformation of the ordinary Yang-Mills partition function. In a special case his expression coincides with the Verlinde formula of the conformal quantum filed theory.

Work on the generalization of the Poisson-sigma model to manifolds with boundary, which was already 
initiated by Catteneo and Felder for the case where the world sheet has the topology of a 
two-dimensional disc, is still under progress. Recently, Falceto and Gawedzki have clarified the relation of the boundary version of the gauged WZW model with a 
Poisson-Lie group $G$ as the target space to the topological Poisson sigma model with the dual Poisson-Lie group $G^*$ as the target \cite{GF}.
 
The purpose of the present article is to show that a calculation generalizing that in \cite{HS} leads to an almost closed expression for 
the partition function of the 
Poisson-sigma model on a disc.  Further, by introducing a procedure for glueing manifolds together by identifying certain boundary components 
we are able to determine the partition function of the linear Poisson-sigma model on arbitrary oriented two-dimensional manifolds. 

The paper is structured as follows. Sec. 2 starts with  a brief review of the Poisson-sigma model, 
including the gauge-fixed extended action of the Batalin-Vilkovisky quantization scheme.  We then perform the calculation 
of the partition function on the disc. We show that for a linear Poisson structure on $\mathds{R}^3$ 
 the partition function of the $SU(2)$ Yang-Mills theory on a disc is recovered. In Sec. 3 we introduce a glueing prescription 
and evaluate the partition function for the linear Poisson-sigma model on arbitrary base manifolds. 
Finally, Sec. 4 contains the conclusions and an outlook for further research.
%
%
%%%%%%%%%%%%%%%%%%%%%%%%%%%%%%%%%%%%%%%%%%%%%%%%%%%%%%%%%%%%%%%%%%%%%%%%%%%%%%%%%%%%%%%%%
%
\section{The partition function on the disc}
%
%%%%%%%%%%%%%%%%%%%%%%%%%%%%%%%%%%%%%%%%%%%%%%%%%%%%%%%%%%%%%%%%%%%%%%%%%%%%%%%%%%%%%%%%%
%
The Poisson-sigma model is a semi-topological field theory on a two dimensional world sheet $\Sigma_g$, 
where $g$ denotes the genus of the manifold. The theory involves a set $X^i$ of bosonic scalar fields which 
can be interpreted as a set of maps $X^i:\Sigma_g\rightarrow N$, where $N$ is a Poisson manifold. 
In this article the Poisson manifolds considered are isomorphic to $\mathds{R}^n$. In addition one has a 
one-form $A_i$ on the world sheet taking values on $T^\ast N$. Due the splitting theorem of Weinstein 
\cite{WE} there exist so-called Casimir-Darboux coordinates $X^i\rightarrow (X^I,X^\alpha)$ for the target manifold, with the properties that the $X^I$ are a complete set of Casimir functions, and the $X^\alpha$ are Darboux coordinates on the corresponding leaves.
In these coordinates the action is given as 
\begin{gather}
\mc{S}[A_I,A_\alpha,X^I,X^\alpha]=\int\limits_{\Sigma_g}\left(A_I\wedge{\rm d}X^I+A_\alpha\wedge\rd X^\alpha+
\frac{1}{2}P^{\alpha\beta}(X^I)A_\alpha\wedge A_\beta+\mc{C}(X^I)\right)
\end{gather}
with $\mc{C}(X)=\mu C(X)$, where $\mu$ is the volume form of the world sheet and $C(X)$ a Casimir function 
of the Poisson bivector $P$. The world sheet can have a boundary, we take it to have the topology of 
a two dimensional disc,  so we must specify the boundary conditions. Denoting by $u$ the 
coordinates of the world sheet, the fields $A_i$ $(i=(I,\alpha))$ are restricted 
to obey $A_i(u)\cdot v=0$ for $u\in\partial \mathbb{D}^2$ and $v$ a vector tangent to the boundary 
\cite{CF1}. 

Due to the fact that the model possesses gauge invariances which have to be taken into account one must 
modify the action in order to perform the path integral quantization. We used for this purpose the antifield formalism of Batalin and Vilkovsky \cite{BV}. The resulting {\it gauge-fixed extended action} is then \cite{HS}
\begin{multline}
S_{gf}[A_I,A_\alpha,X^I,X^\alpha]=  \int\limits_{\mathbb{D}^2}\left(A_I\wedge{\rm d}X^I+A_\alpha\wedge
\rd X^\alpha+
\frac{1}{2}P^{\alpha\beta}(X^I)A_\alpha\wedge A_\beta+\mc{C}(X^I)\right.\\
\left.+\bar{C}^{ J}\frac{\partial\chi_J(A_J)}{\partial A_I}\wedge{\rm d}C_{I}+
\bar{C}^{\alpha}\frac{\partial\chi_\alpha(X_\alpha)}{\partial X_\beta}P^{\gamma\beta}C_{\gamma}
-\bar{\pi}^I\chi_I(A_I)-\bar{\pi}^\alpha\chi_\alpha(X^\alpha)\right)\;.
\label{action}
\end{multline}
The underlying geometry of the antifield formalism was described in the paper of Alexandrov et.al 
\cite{AKSZ}. In \cite{CF2} this approach was extended to the case of world sheets with boundary and applied to the Poisson-sigma model to calculate the extended 
action. They recovered in this approach the boundary conditions which they used in \cite{CF1}. In that paper it was pointed out that the Hodge dual 
antifields have the same boundary condition as the fields. It then follows for $u\in\partial\mathbb{D}^2$ 
that $C_i(u)=0$ and $A_i(u)\cdot v=0$ for $v$ tangent to the boundary, as well as 
$C^\ast(u)=0$ and $A^\ast(u)\cdot w=0$ for $w$ normal to the boundary. The boundary 
condition for the maps $X^i$ is as follows: one is to include in the path integral only maps which
map all points on the boundary to a single point in the target manifold. 

%
%%%%%%%%%%%%%%%%%%%%%%%%%%%%%%%%%%%%%%%%%%%%%%%%%%%%%%%%%%%%%%%%%%%%%%%%%%%%%%%%%%%%%%%%%%%%
%
\subsection{Calculation of the partition function on a disc}
The partition function for the Poisson-sigma model on a disc is 
\begin{gather}
Z(\mathbb{D}^2,\phi(x))=\int\limits_{\Sigma_\Psi}[DX][DA][D\ldots]\exp(-\frac{i}{\hbar}S_{gf})
\langle\delta_x,\phi(X(u^\partial))\rangle\;,
\label{Ps}
\end{gather}
where $\Sigma_\Psi$ denotes the chosen Lagrangian submanifold associated to the gauge fermion $\Psi$ 
\cite{HS}. $\langle\delta_x,\phi(X)\rangle$ is the Dirac measure, a distribution of order zero. 
$\phi(X(u^\partial))$ is an arbitrary function with support only on the boundary of the disc, 
while $u^\partial$ denotes an arbitrary point on the boundary, and $x$ is a point in the target manifold. 
This distribution ensures the boundary condition for the fields $X$, and reflects the freedom of the fields 
on the boundary of the disc. In general, functions of the 
form $X(u^\partial)$ are observables for the Poisson-sigma model, because of the boundary condition 
$C^i(u^\partial)=0$, $(S,X)\mid_{\partial\mathbb{D}^2}=P^{ij}C_i\mid_{\partial\mathbb{D}^2}=0$, 
as noted in \cite{CF1}.
The complete list of functional integrations which must be performed in Eq. (\ref{Ps}) is given in Ref. \cite{HS}.

If one is interested in submanifolds $S$ of $\mathds{R}^n$ one has to reduce the Dirac measure to 
these submanifolds
\begin{gather}
\langle\delta_x,\phi(x)\rangle\biggl|_S\biggr.=\int_S\omega\phi=:\langle\delta_S,\phi(x)\bigl|_S\rangle\;,
\end{gather}
where $\omega$ is the Leray form, which can be chosen to be proportional to the volume form induced on 
the submanifold by the Euclidian measure on $\mathds{R}^n$ \cite{CD}. Note that the function $\phi$ is restricted 
to the submanifold $S$, and the dependence on the point $x$ passes over to the choice of the specific 
submanifold. If one applies this restriction to the foliation of the Poisson manifold such 
that the symplectic leaves $L$ are the considered submanifolds, the Dirac measure picks the 
symplectic leaf $L$ given by $C(X^I)={\rm const}.$. The form of the partition function in Casimir-Darboux 
coordinates is then
\begin{gather}
Z(\mathbb{D}^2,\phi_L(X^\alpha))=
\int\limits_{\Sigma_\Psi}[D\ldots]\langle\delta_{L},\phi_L(X^\alpha)\rangle
\exp(-\frac{i}{\hbar}S_{gf}).
\end{gather}

All the integrations over the fields may be performed. The calculation is the same as in \cite{HS}.
If one performs the gauge fixing for the $X^\alpha$ the integration over $X^\alpha$ goes over to the sum over the homotopy classes of the maps.
 This has the consequence that the function $\phi_L(X^\alpha)$ does not depend on a specific point 
of the target anymore, but just on the homotopy class $[X^\alpha]$ of the associated map 
$X^\alpha:\mathbb{D}^2\rightarrow L$:
\begin{multline}
Z(\mathbb{D}^2,\phi_\Omega(X^\alpha))= \int\limits_{\Sigma_\Psi}{\rm d}X^I_0 \sum\limits_{[X^\alpha]}
{\rm det}\left(\frac{\partial\chi_\alpha(X^\alpha)}{\partial X^\gamma}P^{\gamma\beta}(X^I_0)\right)_{\Omega^0(M)}
{\rm det}^{-1/2}\left(P^{\alpha\beta}(X^I_0)\right)_{\Omega^1(M)}\\
\times\langle\delta_{L},\phi(X^\alpha)\rangle\exp\left(-\frac{i\hbar}{2}
\int\limits_{\mathbb{D}^2}\Omega_{\alpha\beta}\,{\rm d}X^\alpha\wedge{\rm d}X^\beta\right)\exp\left(-\frac{1}
{\hbar}A_{\mathbb{D}^2} C(X^I_0)\right)\;,
\label{part}
\end{multline}
where the subscript $\Omega^k(M)$ indicates that the determinant results from an integration over $k$-forms 
and $A_{\mathbb{D}^2}$ denotes the surface area of the disc. $\Omega_{\alpha \beta}$ is the symplectic form on the leaf.
All the functional integrations have been performed and one has arrived at an almost closed expression for 
the partition function. The boundary condition is now restricted to a function on the symplectic 
leaves which reflects the freedom of the fields $X$ on the boundary. This means that the boundary condition 
for the fields $X$ is now reduced to each single symplectic leaf characterized by the corresponding 
constant mode $X^I_0$. One can now interpret the boundary condition as follows: the boundary of 
the disc is mapped to a point in the target space and one associates to this point the Leray form of the leaf in which it lies. 

%
%%%%%%%%%%%%%%%%%%%%%%%%%%%%%%%%%%%%%%%%%%%%%%%%%%%%%%%%%%%%%%%%%%%%%%%%%%%%%%%%%%%%%%%%%%%%%%%%
%
\subsection{The linear Poisson structure on $\mathds{R}^3$}
In this section we show that the choice of a linear Poisson structure on the target manifold $\mathds{R}^3$ leads to the partition function 
for $SU(2)$ Yang-Mills theory on the disc. The symplectic leaves are then 
2-spheres $\mathbb{S}^2$, and the Leray form is proportional to the symplectic form on the 
sphere induced by the Poisson structure on $\mathds{R}^3$.

The mappings $X^\alpha:\mathbb{D}^2\rightarrow\mathbb{S}^2$ are characterized by the their degree $n$:
\begin{gather}
\frac{i\hbar}{2}\int\limits_{\mathbb{D}^2}\Omega_{\alpha\beta}\,{\rm d}X^\alpha\wedge{\rm d}X^\beta=
\frac{in\hbar}{2}\int\limits_{\mathbb{S}^2}\omega(X^I_0),
\end{gather}
where $\int \omega(X_0^I)$ is the symplectic volume of the leaf associated to the constant mode $X_0^I$. The sum over the degrees defines a periodic delta-function, such that
\begin{gather}
\sum_n\exp[\frac{in}{2\hbar}\int\limits_{\mathbb{S}^2}\omega(X^I_0)]=\sum_n\delta(\int\limits_{\mathbb{S}^2}
\omega(X^I_0)-\frac{n\hbar}{2})\;.\label{ic}
\end{gather}
This shows that the symplectic leaves must be integral, more precisely 
they are half-integer valued. This connection to the $SU(2)$ Yang-Mills theory was worked out in the canonical formalism by Schaller and Strobl, see \cite{SS}.

If we choose the unitary gauge for the fields $X^\alpha$ then both determinants in the partition function of Eq. (\ref{part}) have 
the same form and it is possible 
to combine them. The number of linearly independent forms on a bordered manifold with 
vanishing tangent components, like the gauge fields and the ghosts, is equal to the relative Betti 
number. It then follows that the combined determinant has as exponent the sum of the Betti numbers, which 
is equal to the Euler characteristic of the world-sheet, where the boundary components are now included. For more 
details see \cite{F}. It follows that the exponent for the disc is just 1. The determinants yield the 
symplectic volume $\Vol(L(X^I_0))$ of the leaf $L(X^I_0)$ \cite{HS}.

The linear Poisson structure gives rise to a Lie algebra structure on the dual space $\mc{G}$. Weinstein \cite{WE} 
has shown that the symplectic leaves are exactly the orbits $\Omega$ of the coadjoint representation of 
the compact connected Lie group $G$ corresponding to the Lie algebra $\mc{G}$. The integrality condition (\ref{ic}) of 
the orbits, respectively the symplectic leaves, reduces them to a countable set $\mc{O}(\Omega(X^I_0))$. 
The final result for the partition function of the {\it linear} Poisson-sigma model on the disc is then
\begin{gather}
Z(\mathbb{D}^2,\phi_{\Omega(X^I_0)})=\sum_{\mc{O}(\Omega(X^I_0))}{\rm Vol}(\Omega(X^I_0))
\chi_{\Omega(X^I_0)}(\phi_{\Omega(X^I_0)})
\exp[A_{\mathbb{D}_2}C(\Omega(X^I_0))]\;,
\label{pf}\end{gather}
where we have introduced the notation $\chi_{\Omega}(\phi_\Omega)=\langle\delta_\Omega,\phi_\Omega(X^\alpha)
\rangle=\int_\Omega\;\omega\phi_\Omega(X^\alpha)$.
 Further, the function $\phi$ is still dependent on the coordinates of the leaf, but due to the fact that one integrates over 
the coadjoint orbit $\Omega(X^I_0)$ with respect to the symplectic form $\omega$, the choice of this
function now depends only on the coadjoint orbits. 

As in the case of closed manifolds \cite{HS}, 
it is possible to identify the partition function of the linear Poisson-sigma model with that of 
the Yang-Mills theory. This is essentially based on the duality of the linear Poisson manifold and 
the Lie algebra. 
In this sense we consider the partition function of Eq. (\ref{pf}) {\it dual} to that of the Yang-Mills theory. To see this one may choose a particular function
\begin{gather}
\phi_{\Omega(X^I_0)}(X^\alpha)=\exp(2\pi i\langle X^\alpha,\bar{X}\rangle)\;,
\end{gather}
where $\bar{X}$ is a point of the dual space, the Lie algebra, and $\langle\cdot,\cdot\rangle$ denotes the {\it duality}
pairing. This distribution is nothing else than the Fourier 
transformation of the measure on the orbits, which is the symplectic structure of the orbit. 
This in turn is related to the character formula of Kirillov \cite{KL}:
\begin{gather}
\chi_\Omega(\exp\bar{X})=\frac{1}{j(\bar{X})}\int\limits_\Omega\exp(2\pi i\langle X^\alpha,\bar{X}\rangle)
\frac{\omega^r}{r!}.
\end{gather}
The key ingredient of the orbit method \cite{KL} is the generalized Fourier transform from the space 
of functions on $G$ to the space of functions on $\mc{G}^\ast$, which is the composition of two maps:

1. The map from functions on $G$ to functions on $\mc{G}$ : $f\mapsto\phi:\phi(\bar{X})=j(\bar{X})
f(\exp\bar{X})$, where $j(\bar{X})=\sqrt{\frac{\rd(\exp\bar{X})}{\rd\bar{X}}}$;

2. The usual Fourier transform, which sends functions on $\mc{G}$ to functions on $\mc{G}^\ast$.

Performing the Fourier transformation explicitly in the present case of $G=SU(2)$ one gets 
\begin{gather}
Z(\mathbb{D}^2,\bar{X})=\sum\limits_{\Omega}\Vol(\Omega)\left(\frac{\sin(4\pi X^I_0\bar{X})}{\bar{X}}
\right)
\exp(A_{\mathbb{D}^2}C(\Omega)).
\end{gather}
The Fourier transformation of the Dirac measure restricted to the 2-spheres is proportional to 
$\sin(4\pi X^I_0\bar{X})/\bar{X}$ \cite{CD}, where $X^I_0$ stands now for the quadratic radius such 
that the argument of the sine function is scaled by the volume of the 2-spheres, which is, by the special case  
$\bar{X}=0$ of Kirillov's character formula, the dimension of the corresponding representation. 
All irreducible representations of 
a compact, connected and simply connected Lie group $G$ correspond to integral coadjoint orbits of maximal 
dimension, in the present case these are the orbits of dimension two, the spheres. The determinant of the 
exponential map $j(\bar{X})$ is for the case of $SU(2)$
\begin{gather}
J(\bar{X})=\frac{\sin(\bar{X})}{\bar{X}}\,.
\end{gather}
This leads to
\begin{gather}
\frac{\sin(4\pi X^I_0\bar{X})}{\bar{X}}*J^{-1}=\frac{\sin(4\pi X^I_0\bar{X})}{\sin(\bar{X})}\;,
\end{gather}
which is exactly the expression for the character of $SU(2)$. The representations are characterized by their 
dimensions $\dim(\lambda)=\Vol(\Omega)=4\pi X^I_0$. Taking into account the symmetrization map 
\cite{KL}, which maps 
the quadratic Casimir $C(\Omega)$ which characterizes the coadjoint orbit into the Casimir 
$C(\lambda)$ of the corresponding representation, one gets for the partition function
\begin{gather}
Z(\mathbb{D}^2,\exp\bar{X})=\sum\limits_{\lambda}{\rm dim}(\lambda)\chi_\lambda(\exp\bar{X})
\exp(A_{\mathbb{D}^2}C(\lambda))\;,
\label{su}\end{gather}
where $\chi_\lambda(\exp\bar{X})$ denotes the character of the irreducible representation $\lambda$ of the $SU(2)$ group.
Equation (\ref{su}) is the partition function of the two-dimensional Yang-Mills theory on a disc \cite{BT}. We see that it is a 
special case of the linear Poisson-sigma model, with 
$\exp(2\pi i\langle X^\alpha,\bar{X}\rangle)$ as the specific function on the boundary, which corresponds to 
 $\exp\bar{X}$ by the identification of the Poisson manifold with its dual, the Lie algebra.
%
%%%%%%%%%%%%%%%%%%%%%%%%%%%%%%%%%%%%%%%%%%%%%%%%%%%%%%%%%%%%%%%%%%%%%%%%%%%%%%%%%%%%%%%%%%%%%%%%%%%%%%%%%%%%%
%
\section{The linear Poisson-sigma model on arbitrary surfaces}
%
%%%%%%%%%%%%%%%%%%%%%%%%%%%%%%%%%%%%%%%%%%%%%%%%%%%%%%%%%%%%%%%%%%%%%%%%%%%%%%%%%%%%%%%%%%%%%%%%%%%%%%%%%%%%%
%
The two-dimensional oriented manifolds are fully classified. Starting with a few standard manifolds it 
is possible to obtain an arbitrary manifold with the help of a glueing prescription \cite{M}. We are 
interested in a glueing prescription for 
the partition function of the Poisson-sigma model, which allows the partition function for the glued manifold to be deduced from the partition functions for the components. We consider the various cases in turn. 
%
%%%%%%%%%%%%%%%%%%%%%%%%%%%%%%%%%%%%%%%%%%%%%%%%%%%%%%%%%%%%%%%%%%%%%%%%%%%%%%%%%%%%%%%%%%%%%%%%%%%%%%%%%%%%%
%
\subsection{$g=0\;,\;n\geq 1$}
First we want to create manifolds with more than one boundary component. Geometrically this means that one
starts with a boundary component, a circle, and deforms it into a rectangle. After that one identifies two 
opposite edges such that an additional 
boundary component is created. A formula which allows one to perform such calculations is provided in Ref. \cite{W}.

For functions $\phi_1,\phi_2\in C(G)$, $G$ a Lie group, define the convolution to be
\begin{gather}
(\phi_1\star\phi_2)(g)=\int_G\phi_1(g^\prime)\phi_2(g^{\prime -1}g)\rd g'\;.
\end{gather}
There exists a well-known equation for the generalized character \cite{WL} :
\begin{gather}
\chi_\lambda\star\chi_{\lambda^\prime}=\frac{\delta_{\lambda\lambda^\prime}}{\dim(\lambda)}\chi_\lambda\;,
\end{gather}
where $\lambda$ denotes an irreducible representation of the group. From this equation, together with the fact that the characters form an orthogonal basis for the central 
functions, one gets
\begin{gather}
\chi_\lambda(\phi_1\star\phi_2)=\frac{1}{\dim(\lambda)}\;\;\chi_\lambda(\phi_1)\chi_\lambda(\phi_2).
\end{gather}
One can shift the group convolution to the Lie algebra with the so-called 
wrapping map, as shown in \cite{DW}. Let $\psi_1,\psi_2$ be $G$-invariant, smooth functions on $\mc{G}$. We denote by $\psi^\wedge$  the Fourier transform to the dual space of $\mc{G}$. Then, since 
$\int_{\Omega_\lambda}\rd\mu=\dim(\lambda)$, 
\begin{gather}
\int\limits_\Omega\psi_1^\wedge\rd\mu\int\limits_\Omega\psi_2^\wedge\rd\mu=\Vol(\Omega)
\int\limits_\Omega\psi_1^\wedge\psi_2^\wedge\rd\mu\,,
\end{gather}
where $\rd\mu$ stands for the measure corresponding to the symplectic form of the coadjoint orbit $\Omega$. 
Translating this into the notation of the previous section one gets
\begin{gather}
\chi_\Omega(\phi_1)\chi_\Omega(\phi_2)=\Vol(\Omega)\chi_\Omega(\phi_1\phi_2)\;.
\label{ident}\end{gather}

Using the formula (\ref{ident}) in Eq. (\ref{pf}) yields
\begin{align}
Z(\mathbb{C},\phi_1^{},\phi_2^{})&=\sum_{\mc{O}(\Omega)}{\rm Vol}(\Omega)\chi_{\Omega}(\phi_1^{}\phi_2^{})
\exp(A_{\mathbb{C}}C(\Omega))\notag\\
&=\sum_{\mc{O}(\Omega)}\text{Vol}(\Omega)\chi_{\Omega}(\phi_1)\chi_{\Omega}(\phi_2)\frac{1}{\text{Vol}(\Omega)}
\exp(A_{\mathbb{C}}C(\Omega)\notag)\\
&=\sum_{\mc{O}(\Omega)}\chi_{\Omega}(\phi_1)\chi_{\Omega}(\phi_2)\exp(A_{\mathbb{C}}C(\Omega))\,.
\end{align}
The result is a partition function containing two functions, one with support on each boundary. Geometrically 
this process can be interpreted as follows. First one deforms the boundary, the circle, into a rectangle such 
that each edge of the rectangle has its own degree of freedom, respectively its own function, on the edge. The freedom on the 
boundary turns into $\chi_\Omega(\phi)=\chi_\Omega(\phi_1\phi_2\phi_3\phi_4)$, where the 
$\phi_i$ denote the corresponding parts of the function $\phi$ with support on the edge $i$ of the 
rectangle. Then one identifies two opposite edges
\begin{gather}
\chi_\Omega(\phi)\rightarrow\chi_\Omega(\phi_1\phi_a\phi_2\phi_a^{-1})=\chi_\Omega(\phi_1\phi_2)\,.
\end{gather}
The resulting surface is of course a {\bf cylinder}. This result can be compared with the results 
achieved in the Dirac quantization scheme by Schaller and Strobl in \cite{SS}. In that paper they performed 
the canonical quantization and solved the operator constraint equation for the linear Poisson structure in 
Casimir-Darboux coordinates. Their result was that the wave functions are restricted to the symplectic leaves, 
as are the functions $\phi$ in our calculation, and hence the distributions $\chi_\Omega(\phi)$. Further, they 
showed that in the general case each integral orbit corresponds to one quantum state. 
This can be seen in our calculation in Eq. (\ref{ic}), the integral condition for the orbits.

Note that by choosing both functions to be $\exp(2\pi i\langle X^\alpha,\bar{X}\rangle)$ one gets the correct
result for the partition function of the two-dimensional Yang-Mills theory on the cylinder \cite{BT}.

The manifold with three boundary components, the next step in our construction, is called the {\bf pants} manifold, and 
its partition function is 
\begin{align}
Z(\mathbb{P},\phi_1,\phi_2,\phi_3)&=\sum\limits_\Omega\chi_\Omega(\phi_1)\chi_\Omega(\phi_a^{}\phi_2^{}
\phi_a^{-1}\phi_3^{})\exp(A_{\mathbb{P}}C(\Omega))\notag\\
& =\sum\limits_\Omega\chi_\Omega(\phi_1)\chi_\Omega(\phi_2\phi_3)\exp(A_{\mathbb{P}}C(\Omega))\notag\\
& =\sum\limits_\Omega\frac{1}{\text{Vol}(\Omega)}\chi_\Omega(\phi_1)\chi_\Omega(\phi_2)\chi_\Omega(\phi_3)
\exp(A_{\mathbb{P}}C(\Omega))\;.
\end{align}
In this way we can get any manifold with an arbitrary number $n\geq 1$ of boundaries; for each boundary 
component there is an additional factor $\chi_\Omega(\phi)$ with the new {\it boundary function}
 $\phi$, 
as well as an additional 
factor $\text{Vol}(\Omega)^{-1}$.
%
%%%%%%%%%%%%%%%%%%%%%%%%%%%%%%%%%%%%%%%%%%%%%%%%%%%%%%%%%%%%%%%%%%%%%%%%%%%%%%%%%%%%%%%%%%%%%%%%%%%%%%%%%%%%%
%
\subsection{$g=0\;,\;n=0$}
We now want to calculate the surface  with genus $g=0$ and no boundary component, which is the 2-sphere. The difference is that now we do not just {\it deform} the manifold as in the previous section, 
here we {\it glue} the manifolds together to get the sphere. For this glueing we define the following 
product  
\begin{gather}
\chi_{\Omega}(\phi)\circledast\chi_{\Omega^\prime}(\phi^{-1}):=\frac{\chi_{\Omega}(\phi)}{\Vol(\Omega)}
\frac{\chi_{\Omega^\prime}(\phi^{-1})}{\Vol(\Omega^\prime)}=\delta_{\Omega\Omega'}\;.
\label{glu}\end{gather} 
By choosing $\phi_1=\phi$ and $\phi_2=\phi^{-1}=1/\phi$ in Eq. (\ref{ident}) one 
gets
\begin{gather}
\chi_\Omega(\phi)\chi_\Omega(\phi^{-1})=\Vol(\Omega)\chi_\Omega(\phi\phi^{-1})=\Vol(\Omega)\chi_\Omega
(\mathds{1})=\Vol(\Omega)^2\;.
\end{gather}
This definition of the glueing product is thus quite natural.

We are now in a position to calculate the partition function for the {\bf sphere} by glueing two discs 
together
\begin{align}
Z(\mathbb{S}^2) &= \; Z(\mathbb{D}^2,\phi)\circledast Z(\mathbb{D}^{\prime2},\phi^{-1})\notag\\
& =\sum_{\mc{O}(\Omega)}\sum_{\mc{O}(\Omega^\prime)}\Vol(\Omega)\Vol(\Omega^\prime)
\underbrace{\left[\chi_{\Omega}(\phi)\circledast\chi_{\Omega^{\prime}}(\phi^{-1})
\right]}_{\delta_{\Omega\Omega^\prime}}
\exp(A_{\mathbb{D}^2}C(\Omega))\exp(A_{\mathbb{D}^{\prime 2}}C(\Omega^\prime)\notag)\\
& =\sum_{\mc{O}(\Omega)}{\Vol}(\Omega)^2\exp\left(A_{\mathbb{S}^2}C(\Omega)\right)\,,
\end{align}
which is exactly the partition function for the linear Poisson-sigma model on the sphere calculated in \cite{HS}.

Another check for the new product is performed by deforming two discs to rectangles, and then glueing two edges together. 
The result should again be a rectangle, i.e. one should obtain the partition function 
for the disc. The partition function takes the form
\begin{align}
Z(\Box,\phi_1\phi_2\phi_3\phi_4)&=
\sum_{\mc{O}(\Omega)}{\rm Vol}(\Omega)\chi_{\Omega}
(\phi_1\phi_2\phi_3\phi_4)\exp(A_{\Box}C(\Omega))\notag\\
&= \sum_{\mc{O}(\Omega)}\chi_{\Omega}(\phi_1\phi_2\phi_3)\chi_\Omega(\phi_4)
\exp(A_{\Box}C(\Omega))
\end{align}
\begin{align}
&\hookrightarrow Z(\Box,\phi_1\phi_2\phi_3\phi_a)\circledast Z(\Box',\phi_a^{-1}\phi_4\phi_5\phi_6)\notag\\ 
&=\sum\limits_\Omega\sum\limits_{\Omega^\prime}\chi_\Omega(\phi_1\phi_2\phi_3)
\bigl[\chi_\Omega(\phi_a)\circledast\chi_{\Omega'}(\phi_a^{-1})\bigr]\chi_{\Omega'}(\phi_4\phi_5\phi_6)\exp(A_{\Box}C(\Omega))\exp(A_{\Box'}C(\Omega))\notag\\
&= \sum\limits_\Omega\sum\limits_{\Omega^\prime}\chi_{\Omega^\prime}(\phi_1\phi_2\phi_3)
\chi_\Omega(\phi_4\phi_5\phi_6)\delta_{\Omega\Omega'}\exp(A_{\Box}C(\Omega))\exp(A_{\Box'}C(\Omega))
\notag\\
&= \sum_\Omega\chi_\Omega(\phi_1\phi_2\phi_3)\chi_\Omega(\phi_4\phi_5\phi_6)\exp(A_{\Box}C(\Omega))\notag\\
&= \sum\limits_\Omega \Vol(\Omega)\chi_\Omega(\phi_1\phi_2\phi_3\phi_4\phi_5\phi_6)\exp(A_{\Box}C(\Omega))\notag\\
&= \sum\limits_\Omega \Vol(\Omega)\chi_\Omega(\phi)\exp(A_{\mathbb{D}^2}C(\Omega))=
Z(\mathbb{D}^2,\phi),
\end{align}
with $\phi=\phi_1\phi_2\phi_3\phi_4\phi_5\phi_6$. This calculation shows that the glueing condition 
(\ref{glu}) is self-consistent.
%
%%%%%%%%%%%%%%%%%%%%%%%%%%%%%%%%%%%%%%%%%%%%%%%%%%%%%%%%%%%%%%%%%%%%%%%%%%%%%%%%%%%%%%%%%%%%%%%%%%%%%%%%%%%%%
%
\subsection{$g= 1\;,\;n\geq 0$}
If one changes the genus of the surface one has to use the glueing product (\ref{glu}). The manifold with 
genus $g=1$ and no boundary is the {\bf torus}. One can get it by glueing together two cylinders:
\begin{align}
Z(\mathbb{T})&=Z(\mathbb{C},\phi_a,\phi_b)\circledast Z(\mathbb{C}^\prime,\phi^{-1}_a,\phi^{-1}_b)\notag \\
&=\sum\limits_{\Omega}\sum\limits_{\Omega^\prime}
\bigl[\chi_\Omega(\phi_a)\circledast\chi_{\Omega'}(\phi_a^{-1})\bigr]
\bigl[\chi_\Omega(\phi_b)\circledast\chi_{\Omega'}(\phi_b^{-1})\bigr]
\exp\left({A_{\mathbb{C}}C(\Omega)}\right)\exp\left({A_{\mathbb{C^\prime}}C(\Omega^\prime)}\right)\notag\\
&=\sum\limits_{\Omega}\sum\limits_{\Omega^\prime}\delta_{\Omega\Omega^\prime}\delta_{\Omega\Omega^\prime}
\exp\left( A_{\mathbb{C}}C(\Omega)+A_{\mathbb{C}^\prime}C(\Omega^\prime)\right)\notag\\
&= \sum\limits_\Omega \exp\left( A_{\mathbb{T}}\,C(\Omega)\right).\label{torus}
\end{align}
The torus is again a manifold without boundary, and one can compare it with the solution in \cite{HS}. 
The Euler character for the torus is zero, so in the partition function  the symplectic volume of the coadjoint orbit
does not appear, and we have the same result as in  \cite{HS}.

The next manifold we consider is the {\bf handle} $\mathbf{\Sigma_{1,1}}$, with genus $g=1$ and 
one boundary component $n=1$. To get this surface one has to take the pants manifold and glue two of the boundary 
components together. Due to the fact that one changes the genus one has to use the glueing product 
(\ref{glu}):
\begin{align}
Z(\Sigma_{1,1},\phi)&=Z(\mathbb{P},\phi,\phi_a,\phi^{-1}_a)\notag\\
 &=\sum_\Omega \frac{1}{\Vol(\Omega)}\chi_\Omega(\phi)\bigl[\chi_\Omega(\phi_a)\circledast
\chi_\Omega(\phi_a^{-1})\bigr]\exp\left(A_{\Sigma_{1,1}}\,C(\Omega)\right)\notag\\
& =\sum_\Omega\frac{1}{\Vol(\Omega)}\chi_\Omega(\phi)\exp\left(A_{\Sigma_{1,1}}\,C(\Omega)\right)
\label{g1n1}\;.
\end{align}

This result enables us to calculate the partition function for the torus in yet a third way. Starting from the handle, we glue a disc onto its boundary. 
\begin{align}
Z(\mathbb{T})&= Z(\Sigma_{1,1},\phi_a)\circledast Z(\mathbb{D}^2,\phi^{-1}_a)\notag\\ 
&=\sum_\Omega\sum_{\Omega^\prime}
\frac{1}{\Vol(\Omega)}\bigl[\chi_\Omega(\phi_a)\circledast\chi_{\Omega'}(\phi_a^{-1})\bigr]
\Vol(\Omega^\prime)\exp\left(A_{\Sigma_{1,1}}\,C(\Omega)\right)\exp\left(A_{\mathbb{D}^2}\,
C(\Omega^\prime)\right)\notag\\
&= \sum_\Omega\frac{\Vol(\Omega^\prime)}{\Vol(\Omega)}\delta_{\Omega\Omega^\prime}
\exp\left(A_{\Sigma_{1,1}}\,C(\Omega)+A_{\mathbb{D}^2}C(\Omega^\prime)\right)\notag\\
&=\sum_\Omega\exp\left(A_{\mathbb{T}}\,C(\Omega)\right)\;,
\end{align}
which is the same result as (\ref{torus}).

By glueing two pants together at two boundaries one gets the manifold $\mathbf{\Sigma_{1,2}}$. 
The resulting partition function is
\begin{align}
Z(\Sigma_{1,2},\phi_1,\phi_2)&=Z(\mathbb{P},\phi_1,\phi_a,\phi_b)\circledast Z(\mathbb{P}^\prime,
\phi_3,\phi^{-1}_a\phi^{-1}_b)\notag\\
 &=\sum_\Omega\sum_{\Omega^\prime}\frac{1}{\Vol(\Omega)}\frac{1}{\Vol(\Omega^\prime)}\chi_\Omega(\phi_1)
\chi_\Omega(\phi_2)\bigl[\chi_\Omega(\phi_a)\circledast\chi_{\Omega'}(\phi_a^{-1})\bigr][\chi_\Omega(\phi_b)
\circledast\chi_{\Omega'}(\phi_b^{-1})\bigr]\notag\\
&\phantom{\sum_\Omega\sum_\Omega\sum_{\Omega^\prime}}\times\exp\left(A_\mathbb{P}\,C(\Omega)\right)
\exp\left(A_\mathbb{P'}\,C(\Omega^\prime)\right)\notag\\
&= \sum_\Omega\frac{1}{\Vol(\Omega)^2}\chi_\Omega(\phi_1)\chi_\Omega(\phi_2)\exp\left(A_{\Sigma_{1,2}}\,
C(\Omega)\right).
\label{12}
\end{align}
Due to the fact that we do not change the genus  we can proceed as in the previous section. Starting  with 
the partition function (\ref{g1n1}) 
\begin{gather*}
Z(\Sigma_{1,1},\phi)  =\sum_\Omega \frac{1}{\Vol(\Omega)}\chi_\Omega(\phi)\exp(A_{\Sigma_{1,1}}C(\Omega)),
\end{gather*}
and applying (\ref{ident}) yields
\begin{align}
Z(\Sigma_{1,2},\phi_1,\phi_2) &= \sum_\Omega \frac{1}{\Vol(\Omega)}\chi_\Omega(\phi_1\phi_2)
\exp(A_{\Sigma_{1,2}}C(\Omega))\notag\\
&= \sum_\Omega \frac{1}{\Vol(\Omega)^2}\chi_\Omega(\phi_1)\chi_\Omega(\phi_2)\exp(A_{\Sigma_{1,2}}C(\Omega))\notag\;,
\end{align}
which is the same result as in Eq. (\ref{12}). In this way one gets the partition function of any surface 
with genus $g=1$ and an arbitrary number $n$ of boundary components: $\mathbf{\Sigma_{1,n}}$.
%
%%%%%%%%%%%%%%%%%%%%%%%%%%%%%%%%%%%%%%%%%%%%%%%%%%%%%%%%%%%%%%%%%%%%%%%%%%%%%%%%%%%%%%%%%%%%%%%%%%%%%%%%%%%%%
%
\subsection{Arbitrary $g$ and $n$}
With the considerations of the previous sections we are in a position to calculate the partition function 
for the linear Poisson-sigma model on an arbitrary two-dimensional (oriented) manifold. The fundamental 
manifold we start with is the pants manifold $\Sigma_{0,3}$. The question is how one can calculate the 
partition function on a manifold $\Sigma_{g,n}$ with arbitrary $n$ and $g$. Starting with the pants manifold it 
should be possible to increase $g$ and $n$ in an arbitrary way. On the other hand one must have the possibility 
to decrease the number of boundary components to zero. Hence, one has three requirements:
\begin{itemize}
\item The adding of a disc, i.e. decreasing the number of boundary components $n$ by one, results in multiplying the 
partition function by a factor  $\Vol(\Omega)$.
\item The glueing of the pants manifold, i.e increasing the number of boundary components by one, results in a factor 
 $\Vol(\Omega)^{-1}$. This is similar to the application of (\ref{ident}).
\item The glueing of $\Sigma_{1,2}$ increases the number of the genus, while in the partition function 
an additional factor of $\Vol(\Omega)^{-2}$ appears.
\end{itemize}
These considerations lead to the following expression for the partition function of the linear 
Poisson-sigma model on an arbitrary surface $\Sigma_{g,n}$:
\begin{gather}
Z(\Sigma_{g,n},\phi_1,\ldots,\phi_n)=\sum_\Omega\Vol(\Omega)^{2-2g-n}\prod_{i=1}^n\chi_\Omega(\phi_i)
\exp[A_{\Sigma_{g,n}}C(\Omega)].
\end{gather}
One sees that the exponent of the volume of the symplectic leaf is exactly the Euler characteristic for 
a two-dimensional manifold with genus $g$ and $n$ boundary components. This is the result which would be expected by consideration of the determinants in the partition function.
If one now chooses for each function the specific one which leads to the Fourier transformation 
for the symplectic measure of the orbit one reproduces the result for the two-dimensional Yang-Mills 
theory on arbitrary oriented manifolds \cite{BT}.
%
%%%%%%%%%%%%%%%%%%%%%%%%%%%%%%%%%%%%%%%%%%%%%%%%%%%%%%%%%%%%%%%%%%%%%%%%%%%%%%%%%%%%%%%%%
%
\section{Concluding Remarks} 
In this article we have shown that it is possible to calculate the partition function of the linear
Poisson-sigma model on an arbitrary oriented two-dimensional manifold. To achieve this result we started with the partition function on the disc and then defined the glueing product (\ref{glu}) to go 
over to manifolds with arbitrary genus and number of boundary components. The result includes the 
case of closed manifolds, which was calculated in \cite{HS} in another way.  An interesting further step towards the general 
quantization of the Poisson-sigma model would be the calculation of the partition function for more general Poisson structures. 

The connection between the Poisson-sigma model and Kontsevich's star product discovered by Cattaneo and Felder \cite{CF1} remains a topic worthy of further research. The use of the star product in the deformation quantization approach to quantum theory provides new insights in quantum mechanics, see
\cite{HH} and references therein, and quantum field theory \cite{DF}. Up to now one has used in these works essentially the Moyal star product, which is limited to functions defined on symplectic manifolds. To treat gauge theories in this approach it will be necessary to work in the more general context of Poisson manifolds, in which case the Kontsevich construction, which yields the star product as a formal series in $\hbar$, is relevant. Here one may hope to gain information on the perturbative expansion from knowledge of the complete partition function, which we have studied in this paper for a particular case. This case, where one deals with a linear Poisson structure, is particularly interesting because of the close connection which here prevails between the Kontsevich product and the Campbell-Baker-Hausdorff formula of group theory \cite{KT}.  
\bigskip\\
\noindent{\small 
This work was supported in part (T.Schwarzweller) by the {\it Deutsche Forschungsgemeinschaft} in connection with 
the Graduate College for Elementary Particle Physics in Dortmund.}  
%%%%%%%%%%%%%%%%%%%%%%%%%%%%%%%%%%%%%%%%%%%%%%%%%%%%%%%%%%%%%%%%%%%%%%%%%%%%%%%%%%%%%%%%%
%

%

\begin{thebibliography}{235}
%
%
\bibitem{SS} P.\,Schaller, T.\,Strobl, Poisson structure induced (topological) field theories in two dimensions, Mod. Phys. Lett. A {\bf 9} (1994) 3129
%
\bibitem{I}  N.\,Ikeda, Two-dimensional gravity and non-linear gauge theory, Annals Phys. {\bf 235} (1994), 435
%
\bibitem{TS} T.\,Strobl, Gravity in two spacetime dimensions, e-print archive: hep-th/0011240
%
\bibitem{KS} T.\,Kl\"osch, T.\,Strobl,  Classical and quantum gravity in 1+1 dimensions, part I: a unifying approach, Class. Quant. Grav {\bf 13} (1996), 52; 
Classical and quantum gravity in 1+1 dimensions, part II. the universal coverings; Class. Quant. Grav {\bf 13} (1996), 2395, Classical and quantum gravity in 
1+1 dimensions, part III: solutions of arbitrary topology, Class. Quant. Grav {\bf 14} (1997), 1689
%
\bibitem{EKS} M.\,Ertl, W.\,Kummer, T\, Strobl, General two-dimensional supergravity from Poisson superalgebras, JHEP {\bf 1} (2001) 42
%
\bibitem{ASS} A. Y.\,Alekseev, P.\,Schaller, T.\,Strobl, The topological G/G WZW model in the generalized momentum representation, Phys.Rev. {\bf D52} (1995), 7146 
%
\bibitem{SS2}  P.\,Schaller, T.\,Strobl, Poisson-$\sigma$-models: a generalization of 2d gravity-Yang-Mills systems, Talk delivered at 
the conference on Integrable Systems, JINR, Dubna, 18.-21.7.94, e-print-archive: hep-th/9411163
\bibitem{CF1} A.\,Cattaneo, G.\,Felder, A path integral approach to the Kontsevich quantization formula, Commun. Math. Phys {\bf 212} (2000) 591
%
\bibitem{K} M.\,Kontsevich, Deformation quantization of Poisson manifolds I, e-print archive: q-alg/970904
%
\bibitem{KLV1} W.\,Kummer, H.\,Liebl, D.V.\,Vassilevich, Exact path integral quantization in generic 2d dilaton gravity, Nucl. Phys. B {\bf 493} (1997), 491
%
\bibitem{KLV2} W.\,Kummer, H.\,Liebl, D.V.\,Vassilevich, Integrating geometry in general 2d dilaton gravity with matter,  Nucl. Phys. B {\bf 544} (1999), 403
%
\bibitem{HS} A.\,Hirshfeld, T.\,Schwarzweller, Path integral quantization of the Poisson-sigma model, Ann. Phys. (Leipzig) {\bf 9} (2000) 2 83
%
\bibitem{KC} C.\,Klimcik, The formulae of Kontsevich and Verlinde from the perspective of the Drinfeld double, Commun. Math. Phys. {\bf 217} (2001) 203
%
\bibitem {D} V.G.\;Drinfeld, Quantum groups, Proceedings ICM, Berkeley (1986)
%
\bibitem{GF} F.\,Falceto, K.\,Gawedzki,  Boundary G/G WZW model and topological Poisson-Lie model, e-print archive:  hep-th/0108206
%
\bibitem{WE} A.\,Weinstein,  The local structure of Poisson manifolds, J. Differential Geometry {\bf 18} (1983) 523 
%
\bibitem{BV} I.A.\,Batalin, G.A.\,Vilkovisky, Gauge algebra and quantization, Phys.Lett. {\bf 69} B (1977) 309
%
\bibitem{AKSZ} M.\,Alexandrov, M.\,Kontsevich, A.\,Schwarz, O.\,Zaboronsky, The geometry of the master equation and topological quantum field theory, 
Int. J. Mod. Phys. A  {\bf 12} (1997) 1405
%
\bibitem{CF2} A.\,Cattaneo, G.\,Felder, On the AKSZ formalism of the Poisson-sigma model, Lett. Math. Phys. {\bf 56} (2001), 163
%
\bibitem{CD} Y.\,Choquet-Bruhat, C.\,Dewitt-Morette with M.\,Dillard-Bleick, Analysis, manifolds and physics, North-Holland Publishing Company 1977, revised 
edition 1982
%
\bibitem{F}  T.\,Frankel, The geometry of physics, Cambridge University Press 1997
%
\bibitem{KL} A.\,Kirillov, Merits and demerits of the orbit method, Bull. Am. Math. Soc., New Ser. {\bf36} No.4 (1999), 4333
%
\bibitem{BT} M.\,Blau, G.\,Thompson, Quantum Yang-Mills theory on arbitrary surfaces, Int. J. Mod. Phys. 
A {\bf 7} (1992) 3781
%
\bibitem{M} W.S.\, Massey,  Algebraic topology: an introduction, Springer-Verlag 1967
%
\bibitem{W} N.J.\,Wildberger, On a relationship between adjoint orbits and conjugacy classes of a Lie group, Canad. Bull. Vol {\bf 33} (3), 1990 
%
\bibitem{WL} A.\,Weil, L'integration dans les groups topologies et ses application, Actualites Sci. et. Ind. {\bf 869}, {\bf 1145} (1941), Hermann and Cie, Paris
%
\bibitem{DW} A.H.\,Dooley, N.J.\,Wildberger, Harmonic analysis and the global exponential map for compact Lie groups, Functional Anal. Appl. 27, No. 1 (1993) 21   
%
\bibitem{HH} P.\,Henselder, A.\,Hirshfeld, Deformation quantization in the teaching of quantum mechanics   (accepted for publication in Amer. Jour. Phys.)
%
\bibitem{DF} M.\,D\"{u}tsch and K.\,Fredenhagen, Perturbative algebraic field theory and deformation quantization, hep-th/0101079
%
\bibitem{KT} V.\,Kathotia, Kontsevich's universal formula for deformation quantization and the Campbell-Baker-Hausdorff formula, I, e-print archive: math.QA/9811174
%
\end{thebibliography}
\end{document}